\documentclass[11pt]{article}

\usepackage{amsfonts}
\usepackage{epsfig}
\usepackage{latexsym}

\def\bequ{\begin{equation}}
\def\eequ{\end{equation}}
\def\barr{\begin{array}}
\def\earr{\end{array}}

\def\ben{\begin{equation}}
\def\een{\end{equation}}
\def\bena{\begin{eqnarray}}
\def\eena{\end{eqnarray}}

\def\spa#1{\phantom{\fbox{\rule[-#1cm]{0cm}{0cm}}}}

\def\b1{e^0}

\newcommand{\be}{\begin{equation}}
\newcommand{\ee}{\end{equation}}
\def\bea{\begin{eqnarray}}
\def\eea{\end{eqnarray}}

\def\be{\begin{equation}}
\def\ee{\end{equation}}
\def\bea{\begin{eqnarray}}
\def\eea{\end{eqnarray}}

\def\lesssim{\mathrel{\hbox{\rlap{\hbox{\lower4pt\hbox{$\sim$}}}\hbox{$<$}}}}
\def\gtrsim{\mathrel{\hbox{\rlap{\hbox{\lower4pt\hbox{$\sim$}}}\hbox{$>$}}}}

\begin{document}
\title{{\huge \bf{Casimir energy and a \\ cosmological bounce}}}
\author{{Carlos A. R. Herdeiro and Marco Sampaio}
\\
\\ {\em Departamento de F\'\i sica e Centro de F\'\i sica do Porto}
\\ {\em Faculdade de Ci\^encias da
Universidade do Porto}
\\ {\em Rua do Campo Alegre, 687,  4169-007 Porto, Portugal}}

\date{October 2005}       
 \maketitle

\begin{abstract}
We review different computation methods for the renormalised energy momentum tensor of a quantised scalar field in an Einstein Static Universe. For the extensively studied conformally coupled case we check their equivalence; for different couplings we discuss violation of different energy conditions. In particular, there is a family of masses and couplings which violate the weak and strong energy conditions but do not lead to spacelike propagation. Amongst these cases is that of a minimally coupled massless scalar field with no potential. We also point out a particular coupling for which a massless scalar field has vanishing renormalised energy momentum tensor. We discuss the backreaction problem and in particular the possibility that this Casimir energy could both source a short inflationary epoch and avoid the big bang singularity through a bounce.  
\end{abstract}

\section{Introduction}
The Casimir effect \cite{Casimir} is a manifestation of the vacuum fluctuations of a quantum field. It was first considered in systems with boundaries, and it is known that the Casimir force is highly sensitive to the size, geometry and topology of such boundaries. In particular it may change from attractive to repulsive depending on such shape \cite{review}. But the Casimir force is also present in systems with no boundaries and a compact topology, since the latter imposes periodicity conditions which resemble boundary conditions.

If our universe is either open or flat, with non-trivial topology, or closed, every quantum field living on it should generate a Casimir type force, which led many authors to study the Casimir effect in FRW models (see \cite{review} and references therein for a review). In the case of a spherical universe, most computations of the Casimir energy, or more generically, of the renormalised stress energy tensor, have focused on conformally coupled scalar fields. For instance, a massless conformally coupled scalar field, the electromagnetic field and the massless Dirac field on an Einstein static universe, have been considered in  \cite{Ford1,Ford2}; their Casimir energies have been shown to be of the form $\alpha/R^4$, with $\alpha$, respectively, $1/(480\pi^2)$, $11/(240\pi^2)$ and $17/(1920\pi^2)$. Note that all of these are positive. Since these are all conformally coupled fields, they have equation of state $p=\rho/3$, and obey the strong energy condition. This means the Casimir force is attractive. But this is not always so in the cosmological context. Zel'dovich and Starobinskii \cite{zeldovich} have indeed verified long ago that the Casimir energy of a scalar field could drive inflation in a flat universe with toroidal topology.

The purpose of this letter is to exhibit a family of quantum scalar fields which originate a repulsive Casimir force in a closed universe, since they violate the strong energy condition. Interestingly, this family includes the simplest case one could consider: a minimally coupled massless scalar field with no potential. Our computation will  be performed in the Einstein Static Universe (ESU), which can be faced as an approximation to a dynamical FRW model in sufficiently small time intervals, and avoids having to deal with the complexities of quantum field theory in a time dependent spacetime, like particle creation.\footnote{Recent work on the Casimir effect on the ESU can be found in \cite{Elizalde:2003ke} and references therein.}

\section{Quantum scalar field with arbitrary coupling in the ESU} 
We consider the ESU, which is a well known solution of the Einstein equations sourced by a perfect fluid with positive energy density ($\rho>0$) and zero pressure ($p=0$) together with a positive cosmological constant ($\Lambda>0$):
$G_{\mu \nu}+\Lambda g_{\mu \nu}
=T_{\mu \nu}$, with $T_{\mu \nu}=\rho u_{\mu}u_{\nu}$.
The metric is 
\[
ds^2_{ESU}=-dt^2+R^2d\Omega_{S^3}=-dt^2+\frac{R^2}{4}\left((\sigma_R^1)^2+(\sigma_R^2)^2+(\sigma_R^3)^2\right) \ . \]
We have written the metric on the unit 3-sphere, $d\Omega_{S^3}$, in terms of right forms on $SU(2)$. In order to achieve a static solution, the cosmological constant and the energy density are related by $\Lambda=\rho/2=1/R^2$. Another viewpoint is that the ESU is supported by a perfect fluid with $p=-\rho/3=-1/R^2$.

It is well known that this universe is unstable against small radial perturbations as was first argued by de Sitter. The reason is that the energy density of the perfect fluid increases/decreases with decreasing/increasing radius, whereas the one of the cosmological constant is kept constant. Since the former gives an attractive contribution and the latter a repulsive one, any displacement from the original equilibrium position will grow, rendering such original position as unstable. But even without such classical perturbations this universe is unstable due to quantum mechanics. These are the instabilities we will focus on, in the spirit discussed at the end of the introduction.

Let us consider a free (i.e with no potential) scalar field $\Phi$, with mass $\mu$ and with a coupling (not necessarily conformal) to the Ricci scalar of the background $\mathcal{R}=6/R^2$, governed by
\[ \left(\Box -\xi \mathcal{R}\right)\Phi=\mu^2\Phi \ . \]
Conformal coupling is obtained in four spacetime dimensions by taking the coefficient $\xi=1/6$ (and the theory is then conformal if $\mu=0$), whereas minimal coupling corresponds to $\xi=0$. The compactness of the spatial sections of the background guarantees a discrete mode spectrum which can be easily obtained using elementary group theory. We take the D'Alembertian in the form
\[\Box=-\frac{\partial^2}{\partial t^2}+\frac{4}{R^2}\left(({\bf k}^R_1)^2+({\bf k}^R_2)^2+({\bf k}^R_3)^2\right)=-\frac{\partial^2}{\partial t^2}+\frac{4}{R^2}{\bf k}^2 \ , \]
where ${\bf k}^R_i$ are the right vector fields dual to $\sigma^i_R$ and ${\bf k}^2$ is one of the two Casimirs in $SO(4)$. Notice that the eigenfunctions of the Klein-Gordon operator $(\Box -\xi \mathcal{R}-\mu^2)$ may be taken in the form
\bequ\Phi_n= e^{-i\omega t}\mathcal{D}_j^{m_L,m_R} \ , \label{eigenfunctions} \eequ
where the index $n$ represents all quantum numbers $j,m_L,m_R$ and $\mathcal{D}_j^{m_L,m_R}$ represents a Wigner D-function \cite{Edmonds}. Such function may be thought of as a spherical harmonic on the 3-sphere or as a matrix element of the rotation operator $\langle j, m_L|\hat{R}(\alpha,\beta,\gamma)| j, m_R\rangle$, 
where $|j,m\rangle$ is the basis of a representation of $SU(2)$ and $(\alpha,\beta,\gamma)$ are Euler angles. It follows straightforwardly that the dispersion relation becomes
\bequ
\omega_j^2=\frac{(2j+1)^2+6\xi-1}{R^2}+\mu^2 \ , \ \ \ \ \ j=0,\frac{1}{2},1,\frac{3}{2},\dots \label{scalaresu} \eequ
with the degeneracy of each frequency being $d_j=(2j+1)^2$, in agreement with the spectrum found in \cite{Ford1}. Note that there are no unstable modes for $\xi\in \mathbb{R}_0^+$, which includes minimal and conformal coupling. This is the range of couplings we will analyse in the following.

Canonical quantisation of the scalar field can be performed unambiguously. One finds the mode expansion
\[
\hat{\Phi}=\sum_n \hat{a}_n^{\dagger}\Psi_n+\hat{a}_n\Psi^*_n \ , \ \ \ \ \ \Psi_n=\sqrt{\frac{2j+1}{2\omega_j V}}\Phi_n \ , \]
with $V=2\pi^2R^3$ being the volume of the constant $t$ hypersurfaces and with the operators $\hat{a}_n^{\dagger},\hat{a}_n$ obeying the usual commutation relation $[\hat{a}_n,\hat{a}_{n'}^{\dagger}]=\delta_{nn'}$.

The classical energy momentum tensor of the scalar field is more conveniently written in the natural tetrad basis ${\bf e}^a=\{dt,R\sigma^1_R/2,R\sigma^2_R/2,R\sigma^3_R/2\}$,
\bequ T_{ab}={\bf k}_a\Phi{\bf k}_b\Phi-\frac{g_{ab}}{2}{\bf k}_c\Phi{\bf k}^c\Phi-\frac{\mu^2}{2}g_{ab}\Phi^2+\xi(G_{ab}-\nabla_a{\bf k}_b+g_{ab}\Box)\Phi^2 \ , \label{emtensor} \eequ
where we have denoted ${\bf k}_a=\{\partial/\partial t, {\bf k}_i^R\}$. 
The conformal case ($\xi=1/6,\mu=0$) has zero trace; quantum mechanically, however, the renormalised energy momentum tensor for a conformally coupled, free massless scalar field  generically develops a trace anomaly, which can be written solely in terms of geometric quantities of the background \cite{Brown,Wald}. In four spacetime dimensions such anomaly takes the form
\bequ\langle T^a_{\ \ a}\rangle_{ren} =\frac{1}{120(4\pi)^2}\left\{C_{abcd}C^{abcd}-\frac{1}{3}(R_{abcd}R^{abcd}-4R_{ab}R^{cd}+4\mathcal{R}^2)\right\} + \chi\nabla^2 \mathcal{R} \ . \label{anomaly} \eequ
The coefficient $\chi$ is renormalisation scheme dependent. This trace is zero for the ESU: the Ricci scalar is constant, the Weyl tensor $C_{abcd}$ vanishes since the geometry is conformally flat and the second Euler density also vanishes, since the Euler characteristic of any odd dimensional sphere is zero. 

\section{Renormalisation}
\begin{figure}
\begin{picture}(0,0)(0,0)
\end{picture}
\centering\epsfig{file=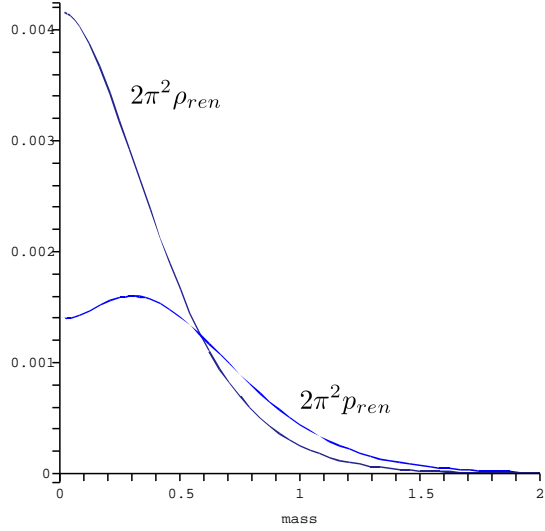,width=8cm}
\caption{Renormalised total vacuum energy $2\pi^2\rho_{ren}$ and pressure $2\pi^2p_{ren}$ for $R=1$, as functions of $\mu\in[0,2]$, for conformally coupled scalar. Both quantities vanish asymptotically with increasing mass.}
\label{rhomassivefig}
\end{figure}

Denoting the non-vanishing components of $\langle T^a_{\ \ b}\rangle$ as $\langle T^0_{\ \ 0}\rangle\equiv-\rho$ and $\langle T^1_{\ \ 1}\rangle=\langle T^2_{\ \ 2}\rangle=\langle T^3_{\ \ 3}\rangle\equiv p$, we find the unrenormalised quantities
\bequ \rho_0=\frac{1}{V}\sum_{n=1}^{+\infty}n^2\frac{\omega_n}{2} \ , \ \ \ 
p_0=\frac{1}{V}\sum_{n=1}^{+\infty}n^2\frac{\omega_n^2-\mu^2}{6\omega_n} \ , \label{massive1} \eequ
with 
\bequ\omega_n=\frac{\sqrt{n^2+\mu^2R^2+6\xi-1}}{R} \ , \ \ \ \ n\in\mathbb{N} \ . \label{wmassive} \eequ
In \cite{Ford2} the particular case with $\xi=1/6$ was studied. These infinite quantities were renormalised by introducing a damping factor $e^{-\beta n}$ in the sums, subtracting the flat space contribution, and taking the parameter $\beta$ to zero. The result is 
\bequ \barr{l} \displaystyle{\rho_{ren}=\frac{1}{4\pi^2R^4}\left\{\sum_{n=1}^{N-1}\left[n^3\sqrt{1+\left(\frac{\mu R}{n}\right)^2}-n^3-\frac{(\mu R)^2}{2}n+\frac{(\mu R)^4}{8n}\right]
+\sum_{n=N}^{+\infty}\sum_{k=3}^{+\infty}b_k\frac{(\mu R)^{2k}}{n^{2k-3}}\right.} \spa{0.6cm}\\ \displaystyle{~~~~~~~~~~~~~~~~~~~~~  \left.+\frac{1}{120}-\frac{(\mu R)^2}{24}-\frac{(\mu R)^4}{32}\left(4\ln{\frac{\mu R}{2}}+1+4\gamma\right)\right\}} \earr \ , \label{rhorenm1}\eequ
\[ \barr{l} \displaystyle{p_{ren}=\frac{1}{12\pi^2R^4}\left\{\sum_{n=1}^{N-1}\left[\frac{n^3}{\sqrt{1+\left(\frac{\mu R}{n}\right)^2}}-n^3+\frac{(\mu R)^2}{2}n-\frac{3(\mu R)^4}{8n}\right]
+\sum_{n=N}^{+\infty}\sum_{k=3}^{+\infty}c_k\frac{(\mu R)^{2k}}{n^{2k-3}}\right.} \spa{0.6cm}\\ \displaystyle{~~~~~~~~~~~~~~~~~~~~~  \left.+\frac{1}{120}+\frac{(\mu R)^2}{24}+\frac{(\mu R)^4}{32}\left(12\ln{\frac{\mu R}{2}}+7+12\gamma\right)\right\}} \earr \ , \]
where $\gamma$ is Euler's constant, $N$ is an integer such that $N>\mu R$ and the explicit formulae for the coefficients $b_k$, $c_k$ are
\[ b_k=\frac{(-1)^{k-1}(2k-3)!!}{2^kk!} \ , \ \ \ \ \ \  c_k=\frac{(-1)^{k}(2k-1)!!}{2^kk!} \ . \]
 These expressions are not very enlightening; therefore we plot the renormalised energy density and pressure, for fixed radius, in terms of the mass in figure \ref{rhomassivefig}. The point we would like to emphasise is that they are always positive. Therefore we conclude that a conformally coupled scalar field with arbitrary mass does not violate the strong energy condition and hence produces solely an attractive effect in the universe.

At this point let us make a comment which also works as a consistency check. The first law of thermodynamics yields $p=-\partial E/\partial V$,  
where the total energy $E=\rho V=\rho 2\pi^2R^3$. We can easily check that this holds both for the infinite unrenormalised quantities (\ref{massive1}) and the finite renormalised ones (\ref{rhorenm1})
\[p_0=-\frac{1}{3R^2}\frac{\partial  (R^3\rho_0)}{\partial R} \ , \ \ \ p_{ren}=-\frac{1}{3R^2}\frac{\partial (R^3\rho_{ren})}{\partial R} \ . \]
Thus, we could have only computed the energy density and have deduced the pressure via the first law of thermodynamics, which obviously only states the conservation of energy $T^{\mu \nu}_{\ \ \ ;\mu}=0$. This is exactly what shall be done in the cases to follow.

As a check on the result (\ref{rhorenm1}) let us consider another renormalisation method, namely zeta function. We can write the regularised expression as
\bequ \bar{\rho}(s)=\frac{\mu_0^{s+1}}{4\pi^2R^{3-s}}\sum_{n=1}^{+\infty}\frac{n^2}{(n^2+a^2)^{s/2}} \ , \label{masszeta} \eequ
where $\mu_0$ has dimension of mass and
\[ a^2\equiv \mu^2R^2+6\xi-1 \ , \]
which, for $a^2\ge 0$, has the appropriate form to be written in terms of Epstein-Hurwitz zeta functions
\bequ  \bar{\rho}(s)=\frac{\mu_0^{s+1}}{4\pi^2R^{3-s}}\left(\zeta_{EH}(s/2-1,a^2)-a^2\zeta_{EH}(s/2,a^2)\right) \ .\label{massreg} \eequ
In order to obtain the renormalised energy density we should now consider the limit of this expression when $s\rightarrow -1$. 

The Epstein-Hurwitz zeta function is only defined by the infinite sum $\sum_{n=1}^{+\infty}(n^2+a^2)^{-s}$ for $Re(s)>1/2$. Its analytic continuation to other values of $s$,  which converges for all $s$ with the exception of an infinite number of poles is given by \cite{nesterenko}
\bequ
\zeta_{EH}(s,a^2)=-\frac{1}{2(a^2)^s}+\frac{\sqrt{\pi}}{2}\frac{\Gamma(s-1/2)}{\Gamma(s)}(a^2)^{1/2-s}+\frac{2\pi^s(a^2)^{1/4-s/2}}{\Gamma(s)}\sum_{n=1}^{+\infty}n^{s-1/2}K_{s-1/2}(2\pi n \sqrt{a^2}) \ , \label{EHzeta} \eequ
where $K_{\nu}$ are modified Bessel functions, and $a^2$ is required to be positive. It is simple to see that the infinite series converges for all $s$, since the modified Bessel functions have asymptotic behaviour
\[
K_{\nu}(z)\sim \sqrt{\frac{\pi}{2z}}e^{-z} \ , \ \ \ \  \ |z|\rightarrow \infty \ . \]
The poles of $\zeta_{EH}(s,a^2)$ arise at the poles of $\Gamma(s-1/2)$, that is at $s=1/2-n$, $n\in \mathbb{N}_0$.

Since $a^2$ must be positive this technique does not apply in the case of a massless minimally coupled scalar field. Let us focus on the case with $a^2>0$. Applying (\ref{EHzeta}) to (\ref{massreg}) we find 
\bequ
\barr{l}
\displaystyle{\bar{\rho}(s)=\frac{\mu_0^{s+1}R^s}{2V}\left(\frac{\sqrt{\pi}}{4}\frac{\Gamma\left(\frac{s-3}{2}\right)}{\Gamma\left(\frac{s}{2}\right)}\sqrt{a^2}^{3-s}+\right.}\spa{0.5cm}\\ \displaystyle{\left. ~~~~~~~~+
\frac{2^{\frac{5-s}{2}}\sqrt{\pi}}{\Gamma\left(\frac{s}{2}-1\right)}\sum_{n=1}^{+\infty}\left[\left(\frac{2\pi n}{\sqrt{a^2}}\right)^{\frac{s-3}{2}}K_{\frac{s-3}{2}}(2\pi n \sqrt{a^2})+\frac{a^2}{2-s}\left(\frac{2\pi n}{\sqrt{a^2}}\right)^{\frac{s-1}{2}}K_{\frac{s-1}{2}}(2\pi n \sqrt{a^2})\right]\right) } \earr\  . \label{zeta}  \eequ
The point $s=-1$ is exactly at one of the poles of the analytic continuation of the Epstein-Hurwitz zeta function. That is, the right hand side of the last formula is an infinite quantity, and it might seem that the zeta function method fails. With the correct interpretation this is not so. Let us consider separately two cases:
\begin{description}
\item[$\bullet$] Conformal coupling ($\xi=1/6$) and arbitrary mass: This case is obtained by substituting $a^2=\mu^2R^2$ in (\ref{zeta}). The infinite contribution, i.e  the term with $\Gamma(-2)$ in the $s=-1$ limit,  is exactly the $R$ independent term. Thus, despite its (infinite) contribution to the Casimir energy it will not contribute to the Casimir \textit{force}. This argument would be enough to neglect it, and to suspect this is the flat space contribution. To confirm this is indeed the case compute the flat space result by taking the infinite radius limit of (\ref{masszeta}):
\[
\lim_{R\rightarrow +\infty}\bar{\rho}(s)=\frac{\mu_0^{s+1}}{4\pi^2}\int_0^{+\infty}dk k^2(k^2+\mu^2)^{-s/2}=\frac{\mu_0^{s+1}}{4\pi^2}\frac{\sqrt{\pi}}{4}\frac{\Gamma\left(\frac{s-3}{2}\right)}{\Gamma\left(\frac{s}{2}\right)}(\mu)^{3-s} \ . \]
This is exactly the infinite term obtained after zeta function regularisation. 
\item[$\bullet$] Nonconformal coupling: In the generic case, the divergent term cannot be identified with the flat space contribution, since it does depend on $R$. However, it turns out that the correct result is still obtained by simply dropping out the divergent terms, as will be shown below. Doing so, the final answer for the renormalised vacuum energy density of a scalar field with generic coupling is 
\bequ
\barr{c}\displaystyle{\rho_{ren}=\frac{1}{4\pi^2R^{4}}\sum_{n=1}^{+\infty}\left(\frac{3(\mu^2R^2+6\xi-1)}{2\pi^2n^2}K_{-2}(2\pi n \sqrt{\mu^2R^2+6\xi-1})\right.} \spa{0.4cm}\\ \displaystyle{\left.~~~~~~~~~~~~~~~~~~~~~~~~~~~~~~ +\frac{(\mu^2R^2+6\xi-1)^{3/2}}{\pi n}K_{-1}(2\pi n \sqrt{\mu^2R^2+6\xi-1})\right)}
\  .  \label{rhorenm2} \earr \eequ
\end{description}
An expression equivalent to (\ref{rhorenm2}) was first derived, for the particular case of $\xi=1/6$, in \cite{Dowker:1976pr}, using a different renormalisation technique. It was argued therein that such renormalised energy momentum tensor could lead to a self consistent ESU solution of the semi-classical Einstein equations. We shall come back to this point in section 4.

The advertised check on our calculation of the Casimir energy for a massive, conformally coupled, scalar in the ESU can be performed graphically; plotting the result (\ref{rhorenm2}) for $\xi=1/6$, we have checked that it perfectly coincides with the previous one (\ref{rhorenm1}), plotted in figure \ref{rhomassivefig}.

The zeta function method confirmed the damping factor method used originally in \cite{Ford2}. But is unfortunately inapplicable to the most general situation when $a^2<0$ which includes the interesting case of a massless minimally coupled scalar field. We will now study this case by a variation of the damping factor method. The essential ingredient will be the Abel-Plana formula which we use in the form \cite{Olver}
\bequ \sum_{m=b}^{+\infty}G(m)=\int_b^{+\infty}G(t)dt+\frac{G(b)}{2}+i\int_0^{+\infty} \frac{G(it+b)-G(-it+b)}{exp{(2\pi t)}-1}dt \ . \label{abelplana} \eequ
We regularise the vacuum energy density in the following way
\bequ \bar{\rho}(\beta)=\frac{1}{4\pi^2}\sum_{n=1}^{+\infty}\frac{1}{R}\frac{n^2}{R^2}\sqrt{\frac{n^2}{R^2}+\frac{\mu^2R^2+6\xi-1}{R^2}}e^{-\beta\Omega(n/R)} \ , \label{regomega} \eequ
where $\Omega$ is a positive function of its argument that grows sufficiently fast with $n$ so as to make the sum converge for any $\beta>0$. A convenient choice obeys
\[
\frac{d\Omega}{d(\frac{n}{R})}=\frac{n^2}{R^2}\sqrt{\frac{n^2}{R^2}+\frac{\mu^2R^2+6\xi-1}{R^2}} \ ; \]
it can be written explicitly as
\[ \barr{l} \displaystyle{\Omega\left(\frac{n}{R}\right)=\frac{1}{8}\left\{2\frac{n}{R}\left(\frac{n^2}{R^2}+\frac{\mu^2R^2+6\xi-1}{R^2}\right)^{3/2} - \frac{\mu^2R^2+6\xi-1}{R^2}\frac{n}{R}\left(\frac{n^2}{R^2}+\frac{\mu^2R^2+6\xi-1}{R^2}\right)^{1/2}\right. } \spa{0.4cm}\\ \displaystyle{ \left. ~~~~~~~~~~ - 
\left(\frac{\mu^2R^2+6\xi-1}{R^2}\right)^2\ln \left(\frac{\frac{n}{R}+\left(\frac{n^2}{R^2}+\frac{\mu^2R^2+6\xi-1}{R^2}\right)^{1/2}}{ \sqrt{\frac{|\mu^2R^2+6\xi-1|}{R^2}}}\right)\right\}} \ . \earr  \]
The integration constant has been chosen such that $\Omega \rightarrow 0$ as the frequency vanishes, which means that the damping factor is not altering the long wavelength modes. This seems to be the most physical choice, since these long wavelength modes are the ones responsible for the Casimir energy. We divide the situation with a stable spectrum into two cases which we analyse separately:

$\bullet$ {$\mu^2R^2+6\xi-1>0$ (includes conformal coupling with arbitrary mass)}. Applying the Abel-Plana formula (\ref{abelplana}) with $b=0$ to the regularised expression (\ref{regomega}) we find
\[ \barr{l} \displaystyle{ \bar{\rho}(\beta)=\frac{1}{4\pi^2}\int_0^{+\infty}d\Omega e^{-\beta \Omega}~~~~~ } \spa{0.4cm}\\ \displaystyle{~~~~ -\frac{i}{4\pi^2R^4}\int_0^{+\infty}t^2\left[\frac{\sqrt{\frac{it-ia}{it+ia}}(it+ia)e^{-\beta\Omega(it/R)}-\sqrt{\frac{-it-ia}{-it+ia}}(-it+ia)e^{-\beta\Omega(-it/R)}}{exp(2\pi t)-1}\right]dt}\ . \earr  \]
The first integral became exactly the contribution of flat space (which justifies our choice of $\Omega$). Indeed, from (\ref{regomega}) 
\bequ \lim_{R\rightarrow +\infty}\bar{\rho}(\beta)=\frac{1}{4\pi^2}\int_0^{+\infty}dx x^2\sqrt{x^2+\mu^2}e^{-\beta\bar{\Omega}(x)}=\frac{1}{4\pi^2}\int_0^{+\infty}d\bar{\Omega} e^{-\beta \bar{\Omega}} \ , \label{flats} \eequ
where 
\[ \bar{\Omega}(x)=\lim_{R\rightarrow \infty}\Omega(x)\ . \] 
The second integral, where $a=\sqrt{\mu^2R^2+6\xi-1}$, converges in the $\beta\rightarrow 0$ limit. Thus
\[ \rho_{ren}=-\frac{i}{4\pi^2R^4}\int_0^{+\infty}t^2\left[\frac{\sqrt{\frac{it-ia}{it+ia}}(it+ia)-\sqrt{\frac{-it-ia}{-it+ia}}(-it+ia)}{exp(2\pi t)-1}\right]dt\ .\]
The square root representation in the complex plane that has been used, has a branch cut for $t\in [-a,a]$. Splitting the $t$ integral in $\int_0^a+\int_a^{+\infty}$, one can check that only the second one contributes. The final result is
\bequ \rho_{ren}=\frac{1}{2\pi^2R^4}\int_{\sqrt{\mu^2R^2+6\xi-1}}^{+\infty}\frac{t^2\sqrt{t^2-(\mu^2R^2+6\xi-1)}}{exp(2\pi t)-1}dt \ . \label{rhorenm3} \eequ
This expression was first obtained for the special case of conformal coupling $\xi=1/6$ in \cite{mamayev} (see also \cite{review}). Specialising for such coupling, the plot of this function exactly coincides with the ones in figure \ref{rhomassivefig}, again checking the result. It is quite striking that expressions apparently as distinct as (\ref{rhorenm1}), (\ref{rhorenm2}) and (\ref{rhorenm3}) are actually different representations of the same function. 

For a more general coupling with $\mu^2R^2+6\xi-1>0$, we can also check that the result (\ref{rhorenm3}) coincides with the one computed with the zeta function method (\ref{rhorenm2}), where we dropped out the divergent term, even though it could not be clearly interpreted as the flat space contribution. Such term should seems to renormalise the gravitational action in the way described in \cite{Streeruwitz:1975sv}. Since we are mostly interested in the renormalised energy-momentum tensor, which will be determined by the finite part of the result, we will not dwell any longer on this point.
\begin{figure}
\begin{picture}(0,0)(0,0)
\end{picture}
\centering\epsfig{file=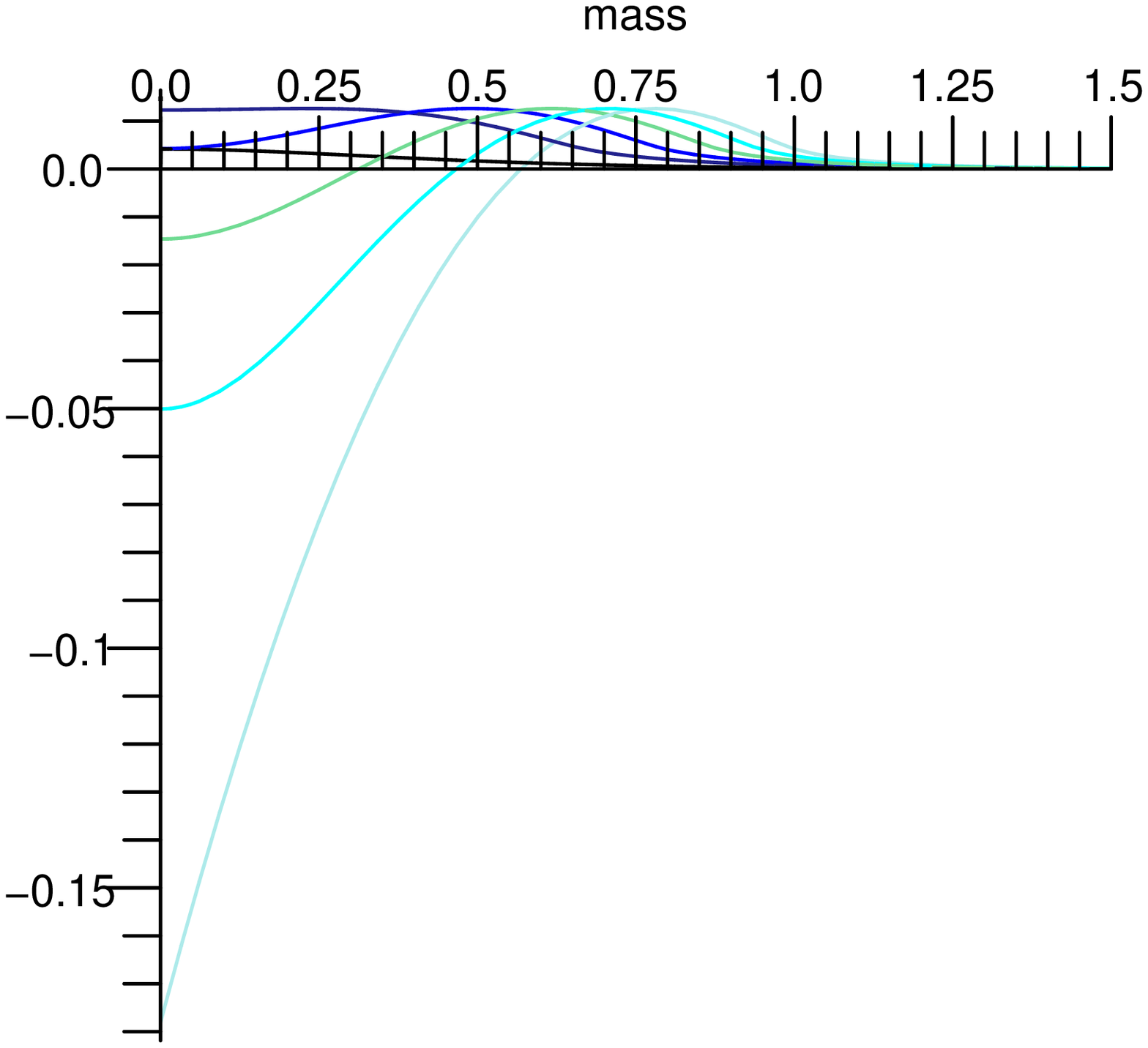,width=8cm}
\centering\epsfig{file=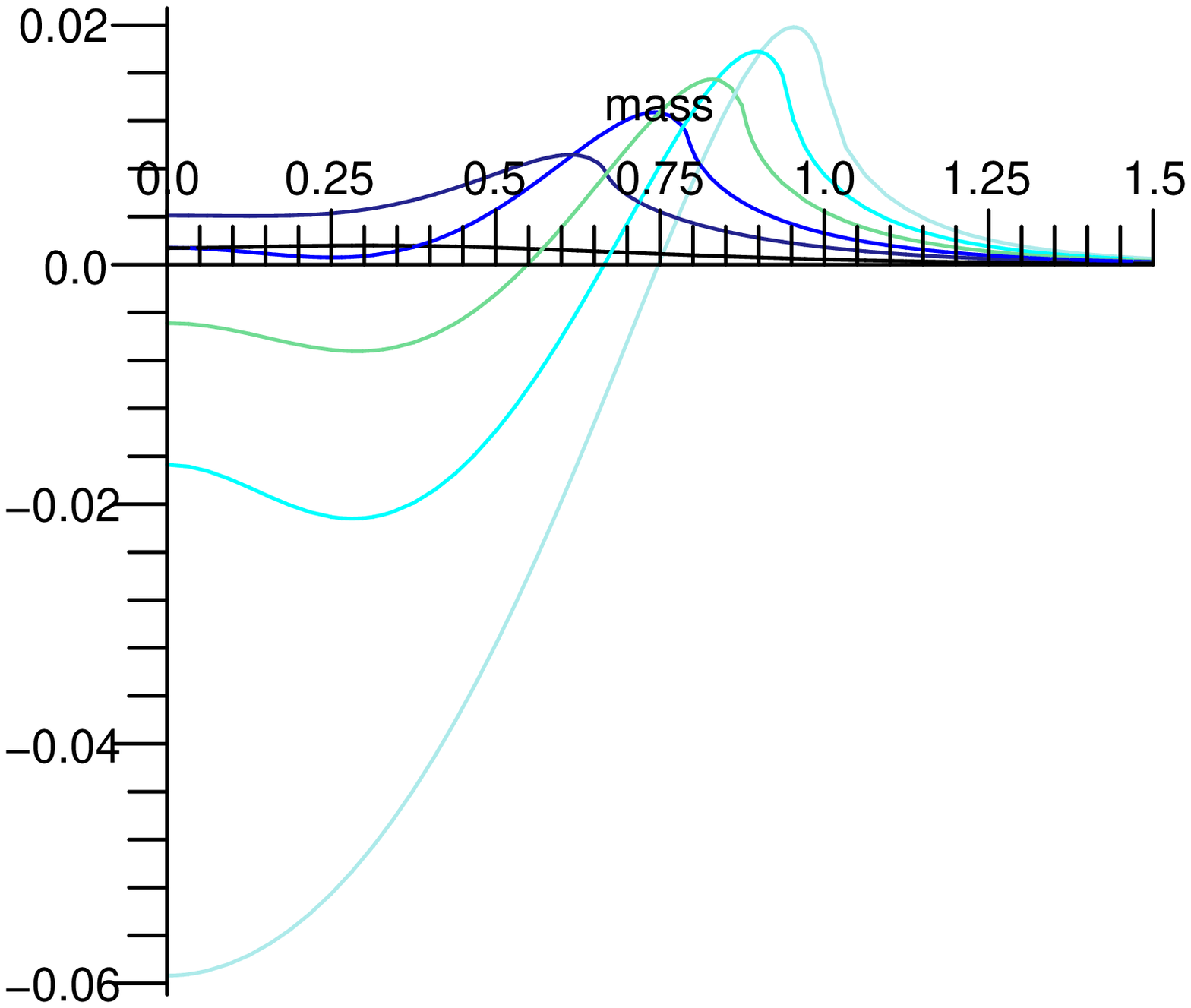,width=8cm}
\caption{Renormalised total vacuum energy $2\pi^2\rho_{ren}$  and pressure  $2\pi^2p_{ren}$ for $R=1$, as a function of $\mu\in[0,1.5]$, for six different couplings $\xi\in [0,1/6]$. As $\xi$ increases the colour of the line in the corresponding graph becomes darker. $\xi=0$ corresponds to the most negative curve in both graphs.}
\label{minimal}
\end{figure} 

\begin{figure}
\begin{picture}(0,0)(0,0)
\end{picture}
\centering\epsfig{file=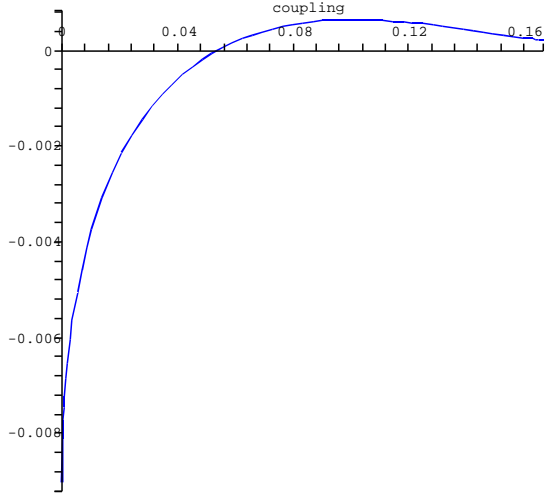,width=8cm}
\caption{Coefficient $\alpha$ in $\rho_{ren}=\alpha/R^4$ and $p_{ren}=\alpha/3R^4$ for a massless scalar as a function of the coupling $\xi$.}
\label{secfig}
\end{figure}

\begin{figure}
\begin{picture}(0,0)(0,0)
\end{picture}
\centering\epsfig{file=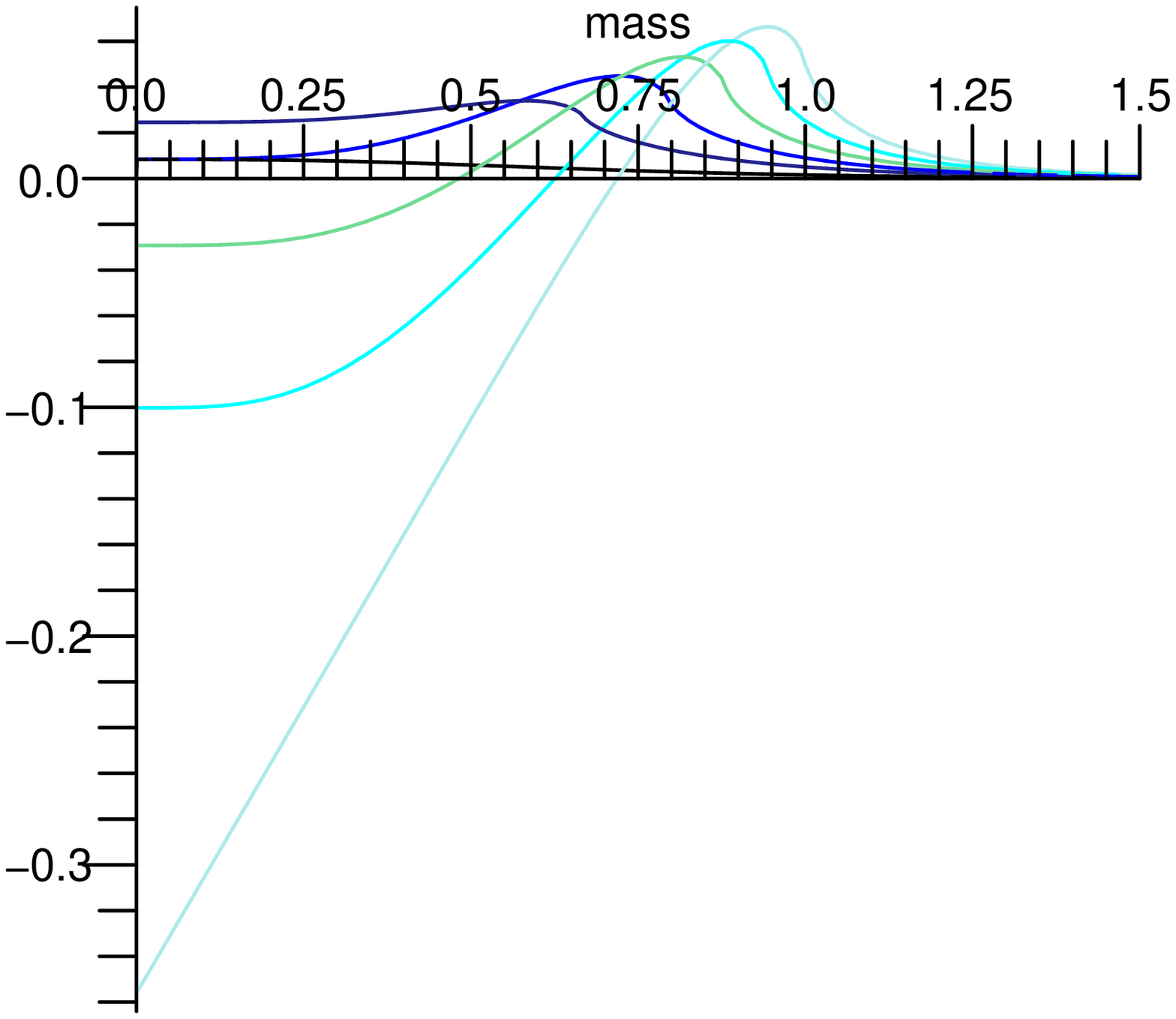,width=8cm}
\centering\epsfig{file=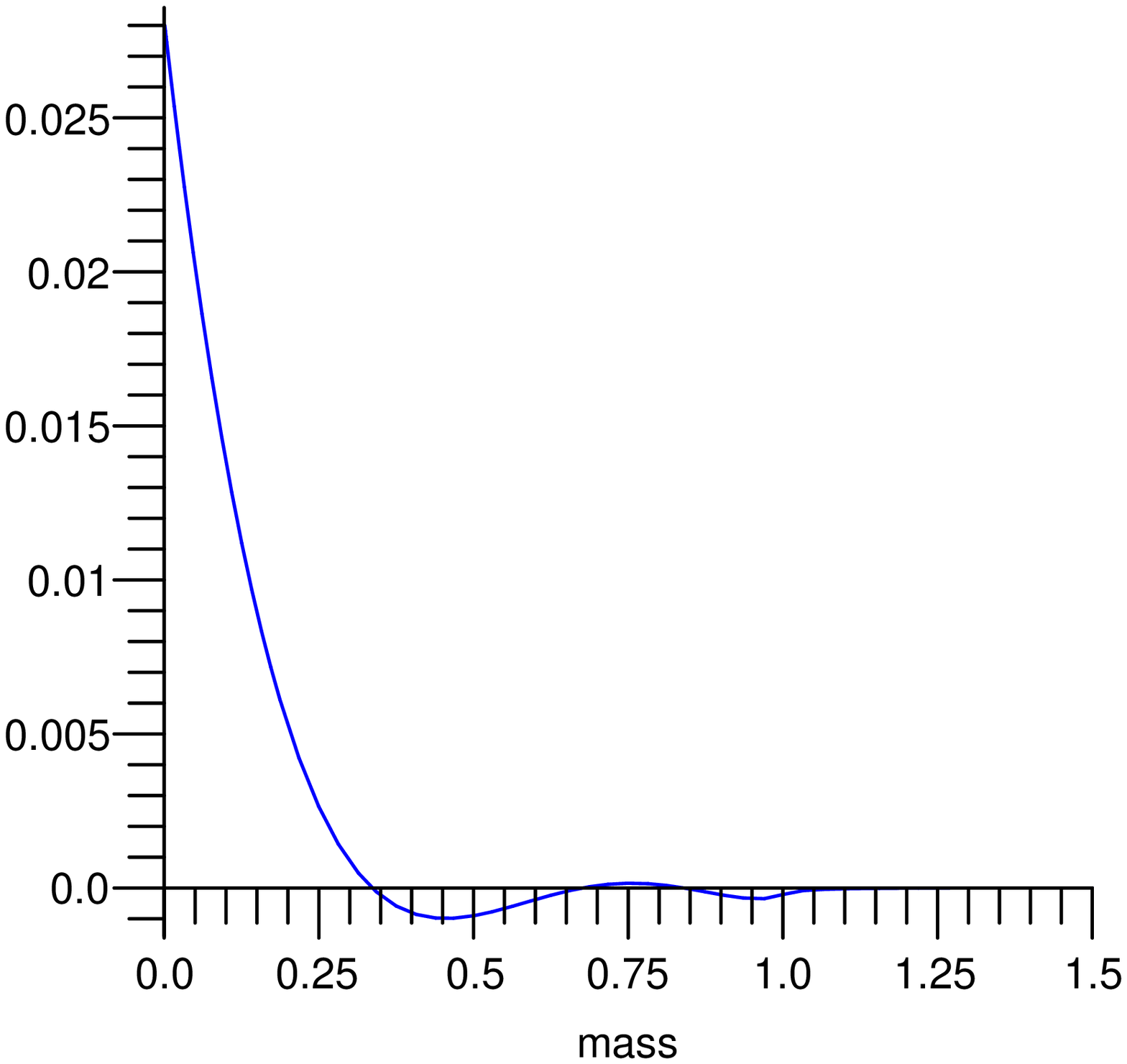,width=8cm}
\caption{Left: $2\pi^2(\rho_{ren}+3p_{ren})$ for six different couplings $\xi\in [0,1/6]$. The colour scheme is the same as in figure (\ref{minimal}). When this quantity is negative the strong energy condition is violated. \mbox{Right: $4\pi^4(\rho_{ren}^2-p_{ren}^2)$} for $\xi=0$; when this quantity is negative the velocity of sound exceeds that of light in the ``quantum fluid'' which implies (but is not necessary for) a violation of the dominant energy condition.}
\label{ec}
\end{figure}

$\bullet$ {$-1\le\mu^2R^2+6\xi-1<0$ (includes minimal coupling with zero mass)}. 
Applying the Abel-Plana formula (\ref{abelplana}) with $b=1$ to the regularised expression (\ref{regomega}) we find
\[ \barr{l} \displaystyle{ \bar{\rho}(\beta)=\int_{\Omega(\frac{1}{R})}^{+\infty}\frac{d\Omega}{4\pi^2} e^{-\beta \Omega}+\frac{\sqrt{\mu^2R^2+6\xi}}{8\pi^2R^4}e^{-\beta\Omega(\frac{1}{R})} +\frac{i}{4\pi^2R^4}\int_0^{+\infty}\frac{f_{\beta}(1+it)-f_{\beta}(1-it)}{exp(2\pi t)-1}dt}\ , \earr  \]
where
\[ f_{\beta}(x)\equiv x^2\sqrt{x^2+\mu^2R^2+6\xi-1}e^{-\beta\Omega(\frac{x}{R})} \ . \]
The flat space contribution is still (\ref{flats}), when $\mu^2>0$. For the tachyonic case the flat space contribution is more subtle, so we will restrict our analysis to $-1\le\mu^2R^2+6\xi-1<0$ and $\mu^2>0$. Subtracting the flat space quantity and removing the regulator $\beta$ we find the renormalised energy density
\bequ \barr{l} \displaystyle{\rho_{ren}=-\frac{1}{32\pi^2R^4}\left[(\mu^2R^2+6\xi)^{\frac{3}{2}}-3\sqrt{\mu^2R^2+6\xi}-(\mu^2R^2+6\xi-1)^2\ln\left(\frac{1+\sqrt{\mu^2R^2+6\xi}}{\sqrt{|1-6\xi-\mu^2R^2|}}\right)\right]} \spa{0.5cm}\\ \displaystyle{~~~~~~~~~ +\frac{i}{4\pi^2R^4}\int_0^{+\infty}\frac{f_{0}(1+it)-f_{0}(1-it)}{exp(2\pi t)-1}dt} \ . \earr \eequ
Despite the apparent unfriendly expression, this is a real quantity which can be plotted without great difficulty. Using the last formula and (\ref{rhorenm3}) we have plotted several cases, with different $\xi$'s, in figure \ref{minimal}. The most noticeable feature is that both the renormalised energy density and pressure may become negative for a range of values. Clearly this leads to violations of the strong energy condition, as will be discussed in more detail in the next section. Let us close by remarking that for zero mass, the renormalised energy density and pressure can be written, for arbitrary coupling, in the form 
\bequ \rho_{ren}=\frac{\alpha}{R^4} \ , \ \ \ \  \ \ \ p_{ren}=\frac{\alpha}{3R^4} \ , \label{masslessprho} \eequ
where 
\bequ\alpha\equiv -\frac{1}{32\pi^2}\left[(6\xi)^{\frac{3}{2}}-3\sqrt{6\xi}-(6\xi-1)^2\ln\left(\frac{1+\sqrt{6\xi}}{\sqrt{|1-6\xi|}}\right)\right]+\frac{i}{4\pi^2}\int_0^{+\infty}\frac{f_{0}(1+it)-f_{0}(1-it)}{exp(2\pi t)-1}dt \ .\eequ
The quantity $\alpha$ is plotted as a function of $\xi$ in figure \ref{secfig}. In particular, the special value of the coupling where $\alpha$ changes sign corresponds to a theory where the renormalised energy momentum tensor vanishes in the ESU. Its numerical value is  $\xi_c\simeq 0.05391$. In figure \ref{ec} some energy conditions are displayed.

\section{Discussion}
\begin{figure}
\begin{picture}(0,0)(0,0)
\end{picture}
\centering\epsfig{file=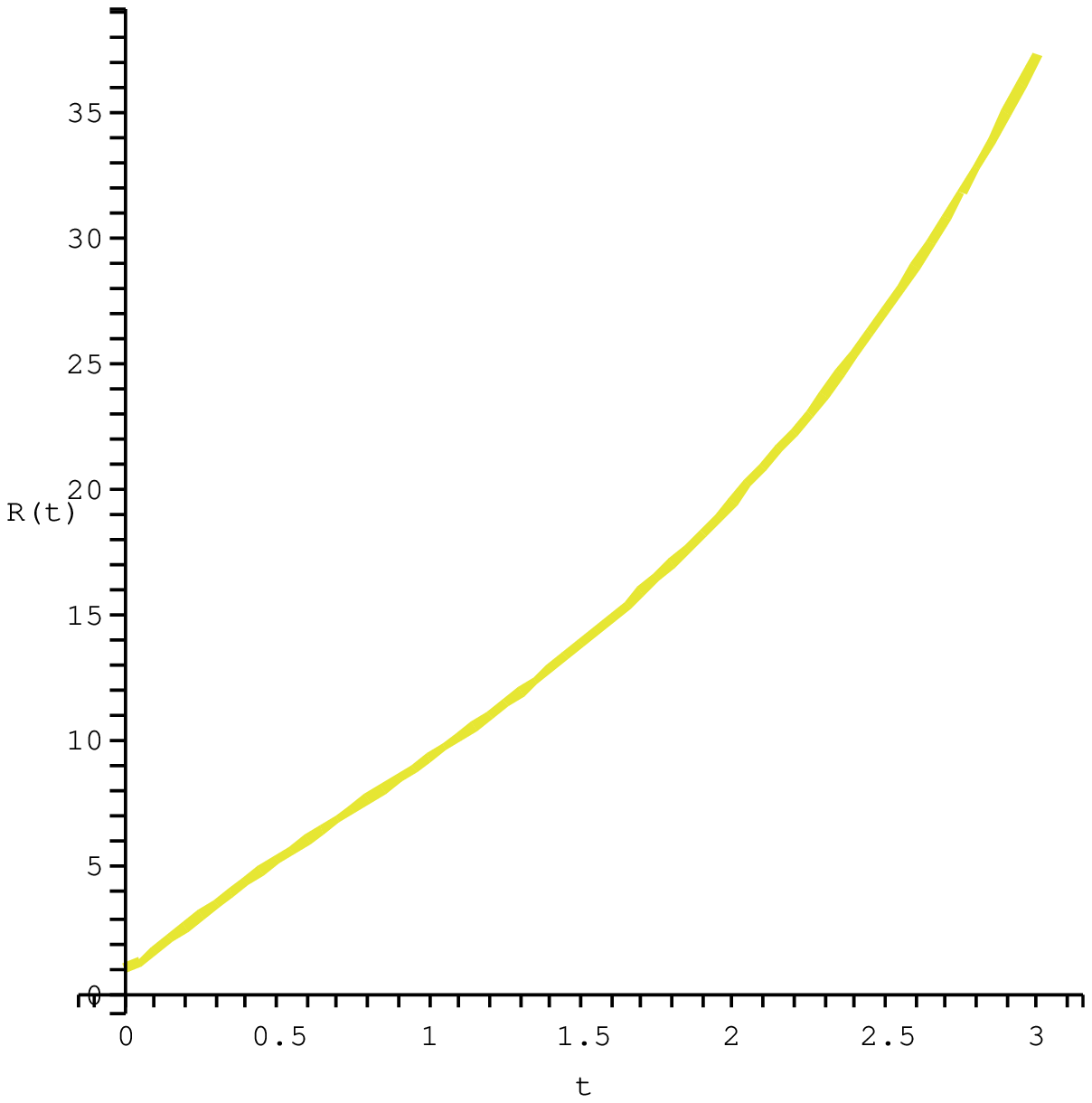,width=8cm}
\centering\epsfig{file=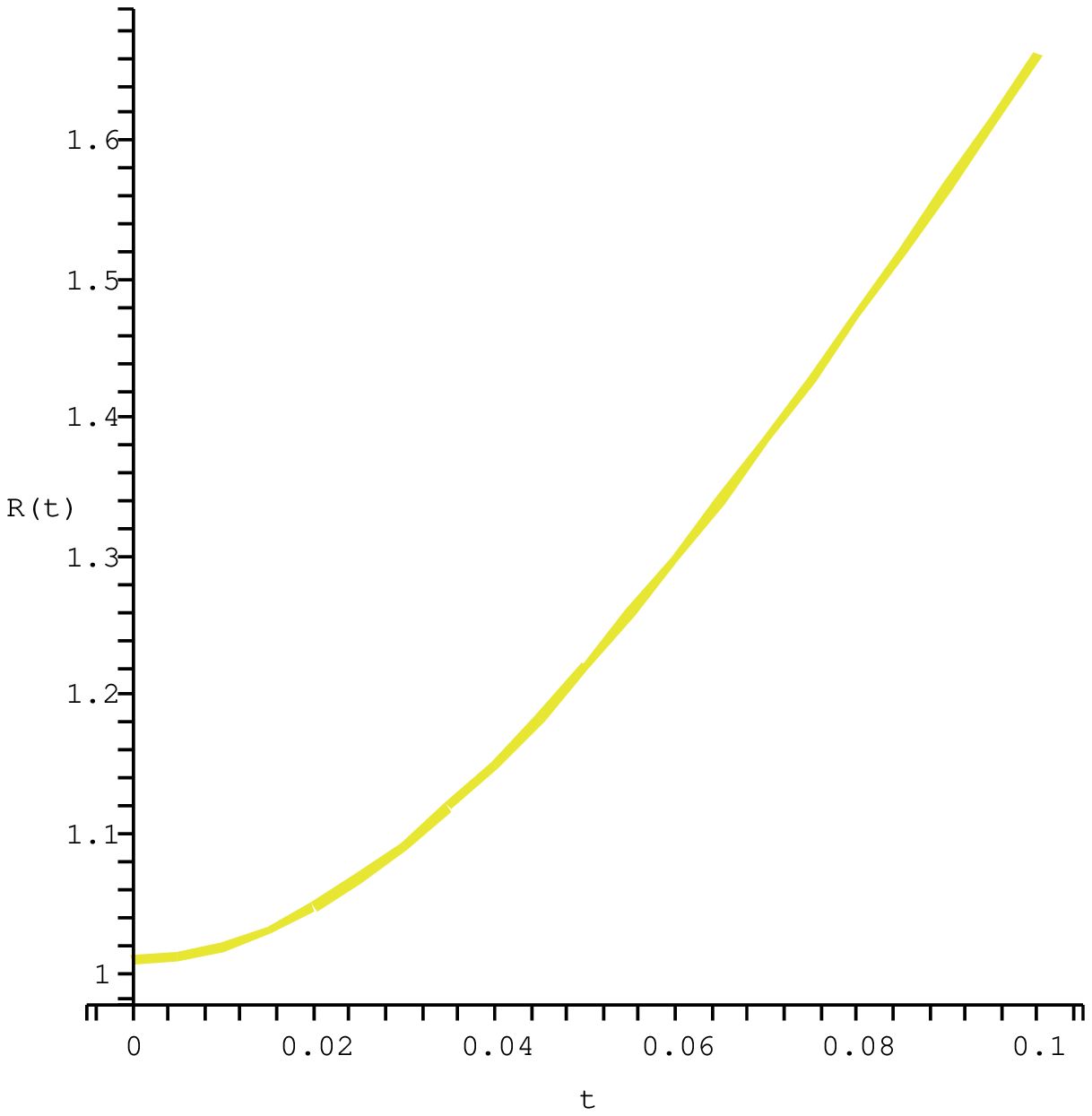,width=8cm}
\caption{Left: scale factor for a universe with a cosmological constant, dust and the quantum fluid of a massless scalar field; Right: Detail near $t=0$ showing clearly the bounce structure.}
\label{sf}
\end{figure}

The main purpose of this paper was to point out that the Casimir effect in the ESU becomes repulsive for a family of scalar fields with various couplings to the Ricci scalar and masses. This can lead to an inflationary era in the early universe, which generically seems to be too short to solve the usual big bang model problems. More interestingly it can lead to a cosmological bounce. To understand how, let us briefly consider the backreaction problem by taking the semiclassical Einstein equations,
\[G_{\mu \nu}+\Lambda g_{\mu \nu}=T^{(matter)}_{\mu \nu}+\langle T^{\phi}_{\mu \nu} \rangle   \ . \]
Analysing the simple massless case (\ref{masslessprho}), for $\alpha>0$, we conclude that the quantum fluid can support a self consistent Einstein Static Universe with some radius, a fact first noticed in \cite{Dowker:1976pr}. Considering the Friedmann equation and Raychaudhuri equations (reinserting Newton's constant) 
\[\dot{R}^2+k=\frac{8\pi G}{3}\rho R^2 \ , \ \ \ \ \ \ddot{R}=-\frac{4\pi G}{3}(\rho + 3p)R \ , \]
and taking the quantum fluid and a positive cosmological constant 
\[ \rho=\Lambda+\frac{\alpha}{R^4} \ , \ \ \ p=-\Lambda+\frac{\alpha}{3R^4} \ , \]
a self consistent solution is obtained with 
\bequ 
k=+1 \ , \ \ \ \ \Lambda=\frac{\alpha}{R^4} \ , \ \ \ \ R=\sqrt{\frac{16\pi G \alpha}{3}} \ . \eequ
Clearly, this solution suffers from fine tuning, and is a universe of the order of the Planck size.
For the case with $\alpha<0$, the fluid can, with the help of dust and a positive cosmological constant, produce an inflationary era followed by a decelerating phase and the present accelerating era. Indeed, taking
\[ \rho=\Lambda-\frac{|\alpha|}{R^4}+\frac{\eta}{R^3} \ , \ \ \ p=-\Lambda-\frac{|\alpha|}{3R^4} \ , \]
one finds the generic behaviour displayed in figure \ref{sf}. The universe has a minimal radius $R=R_c$ where it bounces from a collapsing to an expanding epoch, thus avoiding the Big Bang singularity. For $R>R_c$ there is a short accelerating phase, which is followed by a matter era and then cosmological constant domination.

What is most interesting on having this Casimir energy induced bounce is that it does not rely on any particular form of a potential, and the scalar field needs not having classical effects whatsoever. It is its shear existence that originates the effect. Of course this is by no means a complete cosmological model of our universe. One obvious problem is how to include the radiation era in the picture, which has the same $1/R^4$ dependence for the energy density as the Casimir energy momentum tensor. Indeed, nucleosynthesis data constraints the influence that the quantum fluid could have over that period and therefore, since they vary equally with $R$, over any period where they both exist. A way around this problem could be to assume that radiation is only created after the quantum fluid domination epoch. Such mechanism could be similar to the usual reheating at the end of inflation.

In any case, this simple model illustrates the point that the Casimir energy could both source an early inflationary epoch and avoid the Big Bang singularity in a closed universe. Note that despite having zero mass $\mu$, the non-minimal coupling $\xi$ works, in the Einstein static universe as an effective mass of order $1/R^2$. Thus, this massless case can have its classical dynamics frozen during inflation due to such effective mass. 
. 

It would certainly be interesting to further generalise this analysis to the case of a non-static FRW model where the dynamical Casimir effect takes place. 

\section*{Acknowledgements}
We are very grateful to Pedro Avelino, Filipe Paccetti Correia, Malcolm Perry and Gary Gibbons for discussions and suggestions. We would especially like to thank J.~S.~Dowker for reading the manuscript. C.H. is supported by FCT through the grant SFRH/BPD/5544/2001. M.S. was supported by the Marie Curie research grant MERG-CT-2004-511309. This work was also supported by Funda\c c\~ao Calouste Gulbenkian through \textit{Programa de Est\'\i mulo \`a Investiga\c c\~ao} and by the FCT grants POCTI/FNU/38004/2001 and POCTI/FNU/50161/2003. Centro de F\'\i sica do Porto is partially funded by FCT through POCTI programme.

\end{document}